\documentclass[a4paper,12pt]{article}
\usepackage{epsfig}\parskip 5pt plus 1pt
\usepackage{amsmath}
\usepackage{amssymb}
\usepackage{amsfonts}
\usepackage{fleqn}
\usepackage{mathrsfs}
\usepackage{cite}

\def\be{\begin{equation}}
\def\ee{\end{equation}}
\def\bea{\begin{eqnarray}}
\def\eea{\end{eqnarray}}

\begin{document}

\bibliographystyle{OurBibTeX}

\begin{titlepage}

 \vspace*{-15mm}
\begin{flushright}
\today\\
\end{flushright}
\vspace*{5mm}

\begin{center}
{ \bf \Large Solving the Flavour Problem in Supersymmetric Standard Models with Three Higgs Families}
\\[8mm]
R.~Howl\footnote{E-mail: \texttt{rhowl@soton.ac.uk}.} and
S.F.~King\footnote{E-mail: \texttt{sfk@hep.phys.soton.ac.uk}.}
\\
{\small\it
School of Physics and Astronomy, University of Southampton,\\
Southampton, SO17 1BJ, U.K.\\[2mm]
}
\end{center}
\vspace*{0.75cm}

\begin{abstract}
\noindent We show how a non-Abelian family symmetry $\Delta_{27}$
can be used to solve the flavour problem of supersymmetric standard models
containing three Higgs families such as the Exceptional
Supersymmetric Standard Model (E$_6$SSM). The three 27 dimensional
families of the E$_6$SSM, including the three families of Higgs
fields, transform in a triplet representation of the $\Delta_{27}$
family symmetry, allowing the family symmetry to commute with a
possible high energy E$_6$ symmetry. The $\Delta_{27}$ family
symmetry here provides a high energy understanding of the $Z_2^H$
symmetry of the E$_6$SSM, which solves the flavour changing
neutral current problem of the three families of Higgs fields. The
main phenomenological predictions of the model are tri-bi-maximal
mixing for leptons, two almost degenerate LSPs and two almost
degenerate families of colour triplet D-fermions, providing a
clear prediction for the LHC. In addition the model predicts PGBs with masses
below the TeV scale, and possibly much lighter, which appears to be a
quite general and robust prediction of all models based on
the D-term vacuum alignment mechanism.
\end{abstract}

\end{titlepage}
\newpage

\section{Introduction}
Although TeV scale Supersymmetry (SUSY) is well motivated, the
minimal supersymmetric standard model (MSSM) suffers from the
$\mu$ problem and the little fine-tuning problem
\cite{Chung:2003fi}. The simple replacement of the $\mu$-term of
the MSSM by a singlet superfield coupling to Higgs superfields
leads to a global axial $U(1)$ symmetry whose breaking
would lead to an unwanted axion. In the Next-to-Minimal
Superymmetric Standard Model (NMSSM) a cubic singlet coupling is
also assumed, which breaks the axial $U(1)$ symmetry down to a
discrete $Z_3$ subgroup. In this way the NMSSM solves the $\mu$
problem and the little fine-tuning problem
\cite{BasteroGil:2000bw}. However when the $Z_3$ symmetry of the
NMSSM is broken it leads to cosmological domain walls. One way to
overcome such problems is not to add the cubic singlet term but to
gauge the axial $U(1)$ symmetry so that the would-be axion gets
eaten by a Higgs mechanism resulting in a massive observable $Z'$
\cite{Fayet:1977yc}. Of course such models come at a price since
the gauged $U(1)$ symmetry must be made anomaly-free, and this
generally involves adding additional fermions which are sometimes
ignored in simple phenomenological applications such as in the
USSM \cite{Kalinowski:2008iq}.

The E$_6$SSM \cite{King:2005jy} is a non-minimal supersymmetric
model based on an underlying high-energy E$_6$ symmetry where an
additional low-energy gauged $U(1)$ is identified amongst the
E$_6$ generators, and to cancel anomalies, complete 27 dimensional
E$_6$ families are assumed to survive down to the TeV scale. There
are many possible choices of gauged $U(1)$
\cite{Langacker:2008yv}, but the E$_6$SSM \cite{King:2005jy} is
uniquely defined by a particular choice $U(1)_N$ under which the
right-handed neutrinos are neutral, and so may become heavy,
allowing naturally small physical neutrino masses. As in the
NMSSM, the E$_6$SSM superpotential does not have an explicit
$\mu$-term for the Higgs doublets but instead contains a trilinear
term that represents the interaction between the Higgs doublets
and an additional Standard Model singlet field $S$.  When this
singlet field obtains a vacuum expectation value (VEV) the trilinear term reduces to an
effective $\mu$-term, thus providing a solution to the $\mu$
problem of the MSSM.  The E$_6$SSM also contains an additional
low-energy $U(1)$ gauge symmetry which, as in USSM models, can be
considered to be a gauged Peccei-Quinn symmetry.  This local
$U(1)$ symmetry prevents the presence of an unwanted Goldstone
boson once the singlet field obtains its VEV, which is eaten by
the Higgs mechanism resulting in a massive $Z'$. For this $U(1)$
group to be anomaly free, the entire matter content of the (three
families of) $27$ multiplets of E$_6$ are assumed to survive to
low-energies (apart from the right-handed neutrinos). Each $27$
multiplet contains one generation of quarks and leptons (including
a right-handed neutrino), up and down type Higgs doublet fields (as well as
their coloured partners), and a Standard Model singlet.  The
E$_6$SSM thus contains the particle spectrum of the MSSM plus two
additional MSSM Higgs families, three families of coloured
partners to the three Higgs families, and three singlet fields.
To achieve gauge coupling unification at the GUT scale, the
E$_6$SSM also contains two additional electro-weak doublets from a
$27'$ and $\overline{27'}$ incomplete multiplets of E$_6$ which
are sterile in the E$_6$SSM superpotential. In a minimal version
of the E$_6$SSM, with string scale unification, these additional
states are not present \cite{Howl:2007zi}.

The E$_6$SSM \cite{King:2005jy}, in common with the MSSM and
NMSSM, is subject to the usual flavour problems to do with the
unexplained spectrum of fermion (including neutrino) masses and
mixings of the one hand, and the absence of flavour changing
neutral currents (FCNCs) generically expected in SUSY models on
the other hand. In fact, the E$_6$SSM also faces additional FCNC
challenges due to the three Higgs and singlet families. These
challenges are resolved in the E$_6$SSM by invoking a $Z_2^H$
symmetry which only allows the third family of Higgs and singlet
to couple to quarks and leptons, with the first two Higgs and
singlet families being inert (having zero VEVs) and fermiophobic
(not coupling to quarks and leptons). While this is perfectly
acceptable from the phenomenological point of view, from a
theoretical standpoint the $Z_2^H$ symmetry looks rather {\it ad
hoc} and adds an additional complication to the flavour problem of
the E$_6$SSM not present in the MSSM or NMSSM.

The discovery of neutrino mass and approximately tri-bimaximal lepton mixing \cite{HPS} suggests some kind of a non-Abelian discrete family symmetry might be at work, at least in the lepton sector, and, assuming a GUT-type of structure relating quarks and leptons at a certain high energy scale, within the quark sector too. The observed neutrino flavour symmetry may arise either directly or indirectly from a range of discrete symmetry groups \cite{King:2009ap}. Examples of the direct approach, in which one or more generators of the discrete family symmetry appears in the neutrino flavour group, are typically based on $S_4$ \cite{Lam:2009hn} or a related group such as $A_{4}$ \cite{Ma:2007wu,Altarelli:2006kg} or $PSL(2,7)$ \cite{King:2009mk}. Models of the indirect kind, in which the neutrino flavour symmetry arises accidentally, include also $A_4$ \cite{King:2006np} as well as $\Delta_{27}$ \cite{deMedeirosVarzielas:2006fc} and the continuous flavour symmetries like, e.g., $SO(3)$ \cite{King:2006me} or $SU(3)$ \cite{King:2003rf} which accommodate the discrete groups above as subgroups \cite{deMedeirosVarzielas:2005qg}.
In this Letter we show how a $\Delta_{27}$ family symmetry can resolve all the flavour problems of supersymmetric standard models containing three Higgs families such as the E$_6$SSM \cite{King:2005jy}, including the fermion mass and mixing puzzle present in the SM, the FCNC problems introduced by the MSSM and the additional puzzle of the origin of the $Z_2^H$ symmetry peculiar to models such as the E$_6$SSM. The $\Delta_{27}$ family symmetry is chosen rather than, for example, $A_{4}$ or $S_4$, since it allows complex representations whereas $A_{4}$ and $S_4$ only contain real representations.  Complex representations are required in family symmetry models in which the left-handed matter fields $F$ and right-handed matter fields $F^c$ both transform in triplet representations.  This is to avoid the trivial combination $F F^c h$ where $h$ is the Higgs field. To be concrete we focus on the E$_6$SSM where, under the discrete non-Abelian $\Delta_{27}$ family symmetry, we shall assume that the quarks, leptons and Higgs fields of the E$_6$SSM all transform in triplet representations so that the family symmetry commutes with a possible high energy E$_6$ symmetry. It is known that such a family symmetry can account for various Yukawa couplings responsible for the masses of quarks and leptons, and serves to predict tri-bi-maximal mixing for leptons \cite{deMedeirosVarzielas:2006fc}. It is also known that such family symmetries when applied to supersymmetry can provide a solution to the SUSY flavour and CP problems \cite{Antusch:2008jf}. The qualitatively new feature here is that such a family symmetry can solve the flavour changing neutral current problems introduced by extended Higgs sectors by controlling the Higgs couplings, similar to the $Z^H_2$ symmetry of the E$_6$SSM. As a consequence, we shall find predictions for the mass structure of the three families of Higgs and Higgsino fields and coloured D-fermions of the E$_6$SSM. We remark that in a recent paper \cite{Howl:2008xz} we also considered a non-Abelian $\Delta_{27}$ family symmetry in E$_6$SSM models \cite{King:2005jy,Howl:2007zi}, however the Higgs fields (and D-fermions) were taken to be singlets of the family symmetry, rather than triplets, and the $Z^H_2$ symmetry was simply assumed.

The outline of this Letter is as follows.  In the next section we
describe how the $\Delta_{27}$ family symmetry is applied to the
E$_6$SSM.  This section is split up into subsections which look at
how each term in the E$_6$SSM superpotential is generated from
higher-order operators once we apply the $\Delta_{27}$ symmetry.
In section 2.1 we introduce the E$_6$SSM renormalizable
superpotential, in the absence of any family symmetry.
In section 2.2 we introduce the vacuum alignment required for the various
$\Delta_{27}$ flavon fields.
Then in section 2.3 we describe the
non-renormalizable operators allowed by $\Delta_{27}$ and other symmetries
that lead to the quark and lepton Yukawa
interactions with the Higgs fields.  We also describe the types
of messenger fields that are integrated out to generate the
higher-order operators, and illustrate how the $Z^H_2$ symmetry of
the E$_6$SSM effectively emerges from the high-energy theory.  In section 2.4
we discuss how tri-bi-maximal mixing is generated in our model
from the $\Delta_{27}$ family symmetry and sequential dominance
and then, in section 2.5, we describe how the effective $\mu$-term
of the MSSM is generated and discuss the mass structure of the
LSPs that are formed from the inert higgsinos and singlinos.  In
sections 2.6 and 2.7 the mass structure of the D-fermions is
explained and their decay channels are discussed.
Finally, in section 3, we summarize our findings.



\section{$\Delta_{27}$ Family Symmetry in the E$_6$SSM}

\subsection{The E$_6$SSM Superpotential Without Family Symmetry}

The quark and lepton Yukawa couplings in the E$_6$SSM are derived
from the E$_6$ tensor product $\lambda^{ijk} 27_i 27_j 27_k$ where
$i,j,k$ label the three generations.  To better understand where
the Yukawa couplings come from it is useful to re-write the
superpotential in terms of the Pati-Salam subgroup of E$_6$ under
which each $27$ multiplet decomposes to the following
representations: $ 27 \subset F + F^c + h + D + S$ where $F$
contains the left-handed quark and lepton fields, $F^{c}$ contains
the charge conjugated quark and lepton fields, $h$ contains up and
down Higgs-like fields, $D$ are the colour triplet partners of the
Higgs fields, and $S$ are Standard Model singlets though they are charged under
the gauged $U(1)_N$ symmetry which is broken at low energies \cite{King:2005jy}. In this
Pati-Salam notation the E$_6$ tensor product $27.27.27$ from which
the E$_6$ superpotential is formed, decomposes in the following
way, dropping all couplings and indices for clarity:
\begin{equation} \label{eq:W}
 27. 27. 27 \rightarrow F F^{c} h +  S h h + S D D +  F F D +  F^{c} F^{c}
 D.
\end{equation}
The part of the E$_6$SSM superpotential that contains the Yukawa
interactions for the quarks and leptons is $\lambda^{ijk} F_i
F^{c}_j h_k$ and is the subject of the section 2.3.  In
section 2.5 we discuss the superpotential term  $\lambda_{ijk} S^i
h^j h^k$ from which the MSSM effective $\mu$-term is generated.
section 2.6 describes the term  $\lambda^{ijk} S_i D_j D_k$ from
which the exotic D-fermion states get mass, and section 2.7 looks at the
operators $\lambda^{ijk} F_i F_j D_k + \lambda^{ijk} F^{c}_i
F^{c}_j D_k$ which provide decay channels for the exotic D-fermions.

Although the above operators are written in a Pati-Salam notation
we only do this for ease of notation.  The model presented in this
Letter is in fact based on the Standard Model gauge symmetry
rather than a Pati-Salam symmetry.  For the rest of this Letter we
use a Pati-Salam notation unless stated otherwise.

\begin{table}
\begin{center}
 \begin{tabular}{ | c| c |c |c | c| c| c|}
 \hline
  & $\Delta_{27}$ & $U(1)$ & $Z_2$ & $Z^h_2$ & $Z^S_2$ & $U(1)_R$ \\ \hline
 $F$ & $3$ & 0 & + & + & +&1\\ \hline
 $F^c$ & $3$  & 0 & + & + & +&1\\ \hline
 $h$ & $3$  & 0 & + & - & +&0\\ \hline
 $D$ & $3$  & 0 & + & - & +&0\\ \hline
  $S$ & $3$  & 0 & + & + & -&2 \\ \hline
 $\phi_3$ & $\overline{3}$ & 0 & - & + & + & 0\\ \hline
 $\phi_{23}$ & $\overline{3}$ & -1 & - & + & + & 0\\ \hline
 $\phi_{123}$ & $\overline{3}$ & 1 & - & + & + & 0\\ \hline
  $\overline{\phi}_1$ & $3$ & -4 & - & + & + & 0\\ \hline
  $\overline{\phi}_3$ & $3$ & 3 & + & + & + & 0\\ \hline
 $\overline{\phi}_{123}$ & $3$ & -1 & + & + & +&0\\ \hline
 $\phi^h_3$ & $\overline{3}$ & 0 & + & - & +&0\\ \hline
 $\phi^S_3$ & $\overline{3}$ & 0 & + & + & -&0\\ \hline
 $\overline{\phi}^h_3$ & $3$ & 0 & + & + & +&0\\ \hline
 $\overline{\phi}^S_3$ & $3$ & 0 & + & - & -&0\\ \hline
 $H_{45}$ & 1 & 2 & + & + & +&0\\ \hline
  \end{tabular}
 \caption{\footnotesize This table illustrates how all the flavon fields and Pati-Salam states of the three copies of a $27$ E$_6$ multiplet transform under the $\Delta_{27}$ family symmetry and the additional constraining $U(1) \times   Z_2 \times Z^h_2 \times Z^S_2$ symmetry.  An R-symmetry is also applied to the model which breaks to an R-parity once $S_3$ obtains a vacuum expectation value.}
\end{center}
\end{table}

\subsection{Vacuum Alignment}
In this Letter we add a $\Delta_{27}$ family symmetry
which is broken via the VEVs of the flavon triplets $\phi$ introduced in Table 1.
A number of different triplet flavon fields with different expectation
value directions in $\Delta_{27}$ space are required to produce
the desired mass structure for the E$_6$SSM particles.  Four
different types of directions for the flavon VEVs are used in this Letter
(dropping the VEV brackets):
\begin{equation} \label{eq:VEVs}
\phi_{123} \propto \left( \begin{array}{ccc} 1 & 1 & 1 \end{array} \right),\
\phi_{3} \propto \left( \begin{array}{ccc} 0 & 0 & 1 \end{array} \right),\
\phi_{1} \propto \left( \begin{array}{ccc} 1 & 0 & 0 \end{array} \right),\
\phi_{23} \propto \left( \begin{array}{ccc} 0 & 1 & -1 \end{array} \right).
\end{equation}
In the case of direct models, in which one or more generators of the discrete family symmetry
appears in the neutrino flavour group, the usual mechanism
of vacuum alignment of flavon fields is based on F-term alignment which exploits driving fields in the
superpotential as discussed in \cite{Altarelli:2006kg}.  This mechanism is also available to indirect models, in which
the neutrino flavour symmetry arises accidentally,
as discussed in \cite{deMedeirosVarzielas:2005qg}. However, in the case of indirect models, an additional and elegant
possibility for vacuum alignment becomes available that is not possible for direct models,
namely D-term vacuum alignment introduced in \cite{King:2006np,deMedeirosVarzielas:2006fc}
as discussed below. One advantage of the D-term method is that the terms required
to achieve vacuum alignment originate from the K\"{a}hler potential and so
are not restricted by the symmetries which are introduced to control the terms in the holomorphic
superpotential. Thus the required terms may always be present independent of the symmetries of the model.
In the case of the present model we shall use
the D-term vacuum alignment method to generate the above flavon VEVs, following largely the discussion
in \cite{deMedeirosVarzielas:2006fc,King:2006np}.

An elegant way to obtain the alignments of Eq.\ref{eq:VEVs} is to start with a flavon scalar potential of the form
\cite{King:2006np,deMedeirosVarzielas:2006fc}:
\begin{equation} \label{eq:D-term}
V = m^2 \sum_i \phi^{i\dagger} \phi^i + \lambda \bigg( \sum_i \phi^{i\dagger} \phi^i \bigg)^2 + \Delta V +\ldots,
\end{equation}
where the index $i$ labels the components of a particular flavon triplet $\phi$ and:
\begin{equation} \label{eq:deltaV}
\Delta V = \kappa \sum_i \phi^{i\dagger} \phi^i \phi^{i\dagger} \phi^i.
\end{equation}
In a supersymmetric theory the quartic terms may arise from $D$-terms,
after which this vacuum alignment mechanism is named \cite{King:2006np,deMedeirosVarzielas:2006fc}, which take the form:
\[
\frac{\Big[ \hat{\chi}^{\dagger} \hat{\chi}
(\hat{\phi}^{\dagger} \hat{\phi} \hat{\phi}^{\dagger} \hat{\phi} ) \Big]_D}{M_P^4}
\rightarrow  \frac{F_{\chi}^2}{M_P^4} (\phi^{\dagger} \phi \phi^{\dagger} \phi )
\sim \frac{m_{3/2}^2}{M_P^2}(\phi^{\dagger} \phi \phi^{\dagger} \phi )
\]
where the F-component of the $\Delta_{27}$ singlet $\chi$ acquires a
SUSY breaking VEV $F_{\chi}$, leading to a gravitino mass $m_{3/2}^2 \sim F_{\chi}^2/M_P^2$.
Hence supersymmetry is broken
and the scalar potential $V$ gets a contribution of the type
$(\lambda , \kappa ) \phi^{\dagger} \phi \phi^{\dagger} \phi$
with small $(\lambda , \kappa )\sim {m_{3/2}^2}/{M_P^2}$.
The quadratic mass term originates also from a soft supersymmetry breaking mass term,
and this term is thus expected to have a soft mass squared $m^2 \sim m_{3/2}^2 \sim (TeV)^2$.

We suppose that the flavon mass squared $m^2$ is positive at the (reduced) Planck scale $M_P$
which prevents symmetry breaking at $M_P$ (if $m^2<0$ at $M_P$ then we would expect
a VEV given by $\sqrt{-m^2/\lambda}\sim M_P$).
Then we assume that the soft
mass squared $m^2$ of a given flavon is driven negative by radiative corrections at some scale $\Lambda <M_P$,
which triggers a VEV for that flavon set by the scale $\Lambda$, a mechanism which has been widely used
in different contexts (see for example \cite{Greene:1986jb}).
To see this explicitly here, we may approximate the potential, including radiatively corrected
logarithmically running masses, to be of the form
\footnote{I am indebted to G.Ross (private communication) for this argument.}
$V\approx m^2\phi^{\dagger} \phi \ln (\phi^{\dagger} \phi / \Lambda^2 )$, dropping the small quartic terms.
The first derivative of the potential with respect to $\phi^{\dagger}$
is $V'\approx m^2\phi [\ln (\phi^{\dagger} \phi / \Lambda^2 )+1]$
which is zero for $\langle \phi^{\dagger} \phi \rangle \approx \Lambda^2/e$.
As the different flavons have different
superpotential couplings to heavy states, and since the soft masses run logarithmically with
energy scale, the $\Lambda$ scales defined above may differ greatly for the different flavons. Thus
a hierarchy between the VEVs of various flavon fields is possible, and also stable, in the
framework of the radiative breaking mechanism \cite{Greene:1986jb,deMedeirosVarzielas:2006fc,King:2006np}.

Only the term $\Delta V$ in Eqs.\ref{eq:D-term},\ref{eq:deltaV} determines the alignment,
where $\kappa \sim \kappa_0{m_{3/2}^2}/{M_P^2}$, and the sign of $\kappa_0\sim O(1)$ is undetermined.
For $\kappa > 0$ we obtain
the alignment along the direction of the VEV of
$\phi_{123}$, while $\kappa < 0$ can give rise to that of $\phi_3$.
The configuration $\phi_{23} \propto (0,-1,1)^T$ can then be generated using a
leading higher order term that requires that the VEV is orthogonal to both $(1,0,0)^T$ and $(1,1,1)^T$.
All these operators can be used to generate the VEV configurations of the flavons in Eq.\ref{eq:VEVs}.
A more detailed discussion on this subject can be found in
\cite{deMedeirosVarzielas:2006fc},
where the vacuum alignment operators required for the flavon fields that are used in this Letter
is provided except for the flavon fields $\phi^h_3$,
$\overline{\phi}^h_3$, $\phi^S_3$  and $\overline{\phi}^S_3$.
We assume the same vacuum alignment operators used in
\cite{deMedeirosVarzielas:2006fc} which align all the flavon
fields except for $\phi^h_3$, $\overline{\phi}^h_3$, $\phi^S_3$
and $\overline{\phi}^S_3$.  For these additional flavon fields to
get the required direction of vacuum expectation values, we use
the following D-terms (omitting $m^2_{3/2}/M_P^2 $ factors):
$(\phi^{h}_3)^i
\overline{\phi}_{3 i}  \overline{\phi}^{\dagger}_{3 i}  (\phi^{h
\dagger}_{3})^i$ and $\phi^i_3   \overline{\phi}^h_{3
i}  \overline{\phi}^{h \dagger}_{3 i}  \phi^{ \dagger i}_{3}$
both with  negative coefficients, and similarly for the $\phi^S_3$
and $\overline{\phi}^S_3$ flavons.  These terms cause
$\overline{\phi}^h_3$ and $\phi^h_3$ to get VEVs in the same
direction as the pre-aligned fields $\overline{\phi}_{3 i}$ and
$\phi^i_3$ respectively.


Note that the leading order potential, including $\Delta V$,
is invariant under a product of global $U(1)$
symmetries, one for each flavon component. However this symmetry is broken
explicitly by superpotential terms including the Yukawa superpotential
discussed in the following subsection. It is also broken by higher order
terms in the scalar potential involving purely flavon superfields as indicated by the dots in
Eq.\ref{eq:D-term}.
An example of such a term in the potential which is invariant under all the symmetries in Table~1
but which would violate the additional global $U(1)$ symmetries is
$(m^2_{3/2}/M_P^6 )(\phi_{123}\overline{\phi}_1\overline{\phi}_3)^2(\phi_{23}\phi_{123})^\dag$.
Inserting flavon VEVs of order $\Lambda$,
such a term would lead to pseudo-Goldstone-boson (PGB) square masses of order
$m^2_{PGB} \sim m^2_{3/2}\Lambda^6/M_P^6 $, where generally we expect $\Lambda < M_P$
leading to PGB masses below the TeV scale (and perhaps much lighter) with
possible interesting phenomenological and astrophysical implications
beyond the scope of this paper.
Similarly, inserting the flavon VEVs of order $\Lambda$, one sees that this additional eighth order term
gives a contribution to the potential suppressed relative to the
previous quartic terms by a factor of $\Lambda^4/M_P^4 $ so that, assuming $\Lambda < M_P$,
it will not perturb the previous vacuum alignment arguments appreciably.

It may seem surprising that both the symmetry breaking
and vacuum alignment are governed by quartic terms with very small coefficients
$(\lambda , \kappa )\sim {m_{3/2}^2}/{M_P^2}$. In particular one may worry that
other quartic operators with larger coefficients are present in the theory and that they
could destabilize the symmetry breaking and alignment scheme described above.
However, given the field content and symmetries assumed for the model, in particular the
gauged $U(1)_N$ and the $U(1)_R$ symmetries, it is not possible to write down any operators
which would lead to terms in the flavon potential with competing or larger coefficients
than those of the quartic terms above and
which would destabilize the vacuum, and so the quartic terms
considered above are the dominant ones. On the other hand, we have seen that
such symmetries imply approximate global $U(1)$ symmetries which are only broken
by higher order operators, leading to PGB masses below the TeV scale and possibly much lighter.
The presence of such PGBs seems to be generic to the type of radiative symmetry breaking
associated with D-term vacuum alignment, and this observation
appears to be robust although it has not been remarked upon before.

\subsection{The Effective Yukawa Operators}

Under the $\Delta_{27}$ family symmetry the
$F$, $F^c$ and $h$ transform as triplets so that the
superpotential $\lambda^{ijk} F_i F^{c}_j h_k$ becomes
$\epsilon^{ijk} F_i F^{c}_j h_k$ where $i,j,k$ are now
$\Delta_{27}$ indices. Table 1 describes how all the Pati-Salam
states from a $27$ representation and the flavons transform under
the family symmetry and the additional symmetries that constrain
the model such as $Z^h_2$ symmetry which distinguishes the Higgs
and $D$ fields (but unlike $Z^H_2$ treats all three Higgs families
identically). The lowest order Yukawa superpotential consistent
with the symmetries of Table 1 is:

\[
W_{Yuk} \sim \frac{1}{M^3} F_i F^c_j h_k   \phi^i_{3} \phi^j_{3}
(\phi^{h}_3)^{k}
\]

\[
\hspace{1.1cm} +\frac{1}{M^4} F_i F^c_j h_k  \phi^i_{23}
\phi^j_{23} (\phi^{h}_3)^{k} H_{45}
\]

\[
\hspace{1.1cm}+\frac{1}{M^3} F_i F^c_j h_k   \phi^i_{123}
\phi^j_{23} (\phi^{h}_3)^{k} + \frac{1}{M^3} F_i F^c_j h_k
\phi^j_{123} \phi^i_{23} (\phi^{h}_3)^{k}
\]

\[
\hspace{1.1cm}+\frac{1}{M^6} F_i F^c_j h_k  \phi^i_{123} \phi^j_3
(\phi^m_{123} \overline{\phi}_{1m}) (\phi^{h}_3)^{k} H_{45}
+\frac{1}{M^6} F_i F^c_j h_k  \phi^j_{123} \phi^i_{3}
(\phi^m_{123} \overline{\phi}_{1m}) (\phi^{h}_3)^{k} H_{45}
\]

\begin{equation} \label{eq:1}
\hspace{1.1cm}+\frac{1}{M^7} F_i F^c_j h_k  \phi^i_{123}
\phi^j_{123}
 (\phi^l_{3} \overline{\phi}_{123l}) (\phi^m_{3} \overline{\phi}_{123m})
(\phi^{h}_3)^{k}
\end{equation}
where all order 1 coupling constants are suppressed. $\phi_{23}$,
$\phi_3^h$, $\phi_{123}$, $\phi_3$ are all anti-triplets of the
$\Delta_{27}$ family symmetry, $\overline{\phi}_1$ is a triplet of
$\Delta_{27}$, $H_{45}$ is a singlet of $\Delta_{27}$ but a $45$
of the $SU(5)$ subgroup of E$_6$, and $M$ is some messenger scale
that is discussed further in section 2.3.2.  The  $\phi_3^h$ field is
taken to transform under the $Z^h_2$ symmetry so that all of the
above operators respect this symmetry. We assume that this flavon
field and $\phi_3$ get a VEV in the third component of
$\Delta_{27}$, $\phi_{23}$ gets an equal but opposite VEV in the
second and third components of $\Delta_{27}$, and $\phi_{123}$
gets an equal VEV in the first, second and third components of
$\Delta_{27}$.  The vacuum alignment of these fields was discussed
in section 2.2. For ease of notation we denote $\phi$ as a field
that transforms as $\overline{3}$ under $\Delta_{27}$ and
$\overline{\phi}$ as a field that transforms as a $3$.

In addition to the $\Delta_{27}$ family symmetry, Table 1 also contains the vertical symmetries $U(1) \times Z_2 \times Z^h_2 \times Z^S_2 \times U(1)_R$.  These symmetries prevent interactions that would otherwise be allowed by the $\Delta_{27}$ symmetry and would introduce certain  phenomenological issues to the model.  In particular, The $Z^h_2$ symmetry is used to
differentiate certain superfields including the Higgs from others, and is instrumental in
preventing flavour changing neutral currents due to the extended Higgs sector of the model, as discussed further in the following section.  The $Z^S_2$ is introduced in section 2.5 and is primarily used to lead to
a non-renormalizable operator that reduces to an effective $\mu$-term at the soft SUSY scale, while forbidding other
operators. Note that $Z^S_2$ also forbids the renormalizable operator $\epsilon_{ijk}S^ih_u^jh_d^k$
but this is not a crucial requirement and similar terms are generated at higher order.
The $U(1) \times Z_2$ symmetries are invoked to allow only certain combinations of flavon fields, as in
\cite{deMedeirosVarzielas:2006fc}. For example, the $U(1)$ symmetry prevents the effective Yukawa operator $\frac{1}{M^2} F_i F^c_j \phi^i_{123} \phi^j_3$ from appearing in Eq.\ref{eq:1}. Finally, the $U(1)_R$ symmetry is an R-symmetry that is assumed to reduce to R-parity allowing the LSP of the model to be stable.


\subsubsection{Preventing Flavour Changing Neutral Currents}

As discussed in the Introduction, the E$_6$SSM \cite{King:2005jy}
is subject to the ``usual'' FCNC and CP problems generically expected in SUSY models,
due to SUSY loop diagrams with off-diagonal soft mass insertions,
as well as ``additional'' FCNC challenges arising from tree-level Higgs exchange due to
the three Higgs and singlet families.
The introduction of a family symmetry, together with the idea of spontaneous CP violation by the flavons,
leads to a natural suppression of the ``usual'' FCNC and CP violating
operators and provides a solution to the SUSY flavour and CP problems \cite{Antusch:2008jf}.

The impact of the higher order K\"{a}hler operators has been systematically explored for
$SU(3)$ in \cite{Antusch:2008jf} and the results there are directly applicable to the present
model based on $\Delta_{27}$.
In the exact $\Delta_{27}$ family symmetry limit the soft masses are universal:
\begin{eqnarray}
\hat m_{Q}^2 \propto \hat m_{u^{c}}^2 \propto \hat m_{d^{c}}^2 \propto \hat m_{L}^2
\propto \hat m_{e^{c}}^2 \propto \hat m_{\nu^{c}}^2 \propto \mathbf{1}.
\end{eqnarray}
However, in this limit, the Yukawa couplings and trilinear terms vanish.
In reality the $\Delta_{27}$ family symmetry has to be broken, leading to violations of universality.
For example, the soft masses allowed by the $\Delta_{27}$ family symmetry can be written as:
\[\hat{m}^2_{F,F^c} = m^2_0 \Big( b^{F,F^c}_0  \mathbf{1} + \sum^{F,F^c}_A \frac{\langle \phi_A \phi_A^{\ast} \rangle}{M^m_{F,F^c}} + \cdots \Big)\]
where the generic subscript $A$ runs over all the relevant flavon species. The above off-diagonal
soft mass terms are suppressed by $\epsilon$ factors from the ratios of the flavon VEVs and the messenger masses.  In \cite{Antusch:2008jf} it was rigorously shown that off-diagonal terms from flavons of the type in Table~1 lead to suppressed FCNCs concistent with present experimental limits.
The family symmetry thus naturally suppresses the ``usual'' type of induced FCNCs
due to off-diagonal soft masses and can therefore provide a resolution to the generic SUSY FCNC problem.

In addition to the FCNCs associated with the soft SUSY potential, FCNCs can also originate from
models with the extended Higgs sectors, leading to ``additional'' FCNC problems arising from
tree-level Higgs exchange.
However, as we shall now show,
the $Z^h_2$ symmetry, in combination with the $\Delta_{27}$ family symmetry, suppresses
FCNCs due to tree-level Higgs exchange in the present model.
This can be understood by noting that,
since $\phi^h_3$ transforms under $Z^h_2$, it will
couple to the Higgs fields but not to the quarks and leptons. This
can be understood by considering the messenger diagrams of the
above higher-order operators where $\phi^h_3$ will only be allowed
to attach itself to the Higgs fields (and the Higgs-like messenger
fields) if all the messenger fields are even under $Z^h_2$.  This
is illustrated by Figure~1. Once $\phi^h_3$ gets a VEV, only the
third generation of the up and down Higgs fields $h_3$ are allowed to
couple to the quarks and leptons. It is these up and down Higgs
fields which we therefore take to obtain electro-weak scale VEVs,
and thus act like the up and down Higgs fields of the MSSM.

The $Z^h_2$ and $\Delta_{27}$ symmetries prevent the first and
second generation of Higgs fields from interacting with the quarks
and leptons at tree-level and so there can be no tree-level FCNC
processes involving the neutral scalar components of these fields.
In the E$_6$SSM a $Z_2$ symmetry called $Z^H_2$ is applied to all
the $27$ fields except for the third generation of Higgs fields
and singlet fields to prevent the first and second generation of
Higgs fields from interacting with the quarks and leptons at
tree-level.  The $Z^h_2$ in this Letter is therefore acting as the
$Z^H_2$ symmetry of the E$_6$SSM even though it does not
distinguish between the different Higgs fields.  The $Z^H_2$
symmetry in the E$_6$SSM is broken by an additional discrete $Z_2$
symmetry that forbids the colour triplets of the Higgs fields
causing rapid proton decay \cite{King:2005jy}.  This additional
discrete symmetry will not break the $Z^h_2$ symmetry here, however
any misalignment of $\phi^h_3$ will play the role of $Z^H_2$ breaking,
as discussed later.

\subsubsection{The Messenger Fields}

The messenger fields $\Sigma$ that are responsible for the
suppression factors in Eq.\ref{eq:1} include fields that transform
in the same way as quarks and leptons under the Standard Model
gauge group and as singlets, triplets and anti-triplets of
$\Delta_{27}$.  We refer to these type of messenger field as quark
and lepton-like messengers $\Sigma_{F,F^c}$.  In addition there
are also messengers that are singlets of $\Delta_{27}$ and
transform in the same way as Higgs fields under the Standard Model
gauge group.  We refer to these messenger fields as Higgs-like
messengers $\Sigma_h$.  All these messenger fields are taken to
carry positive $Z^h_2$ parity, and we assume that the Higgs-like
messengers $\Sigma_h$ are heavier than the quark and lepton-like
messengers $\Sigma_{F,F^c}$ so that the latter dominate the
messenger diagrams.  We further assume that the right-handed quark
and lepton messengers $\Sigma_{F^c}$ are heavier than their
left-handed counterparts $\Sigma_{F}$ so that the former dominate
over the latter.  The messenger diagrams are illustrated by Figure
1.

To create a smaller hierarchy in the down quark sector compared to
the up quark sector, we take the mass of the $3$ and
$\overline{3}$ up and down Higgs messengers $M^h_3$ to be equal,
but the up right-handed quark messengers $\Sigma_{u^c}$ that are
$3$ and $\overline{3}$  and singlets of $SU(3)$ to have a mass
$M^{u}$ that is greater than the right-handed down quark
messengers $\Sigma_{d^c}$ by approximately a factor of three. For
the top Yukawa coupling constant to be greater than the bottom
Yukawa coupling constant we take the $\phi_3$ flavon to transform
as a $3 + 1$ of the $SU(2)_R$ subgroup of $E_6$ and choose its VEV
so that $\langle\phi_3\rangle / M_d = \langle\phi_3\rangle / M_u$ as in \cite{deMedeirosVarzielas:2006fc}.
In terms of these messenger masses, the VEV scales for the various
flavon fields are then taken to be the following:
\be
\frac{\langle\phi^h_{3}\rangle}{M^h_3} \approx
\frac{\langle\phi_{3}\rangle}{M^u} \approx 0.8, \ \
\frac{\langle\phi_{23}\rangle}{M^u} \approx \epsilon_u, \ \
\frac{\langle\phi_{123}\rangle}{M^u}
\approx \epsilon^2_u
\ee
where $\epsilon_u \approx 0.05$.
At the GUT scale the Yukawa coupling for the top and bottom quark
is expected to be about $0.5$ in third family Yukawa unification
models based on the MSSM with large $\tan \beta$
\cite{Ross:2007az}.  We therefore assume that $\langle\phi^h_{3}\rangle /
M^h_3 \approx \langle\phi_{3}\rangle / M^u \approx 0.8$. By comparison, in $\Delta_{27}$
models in which the Higgs is a singlet and there is a
messenger-scale suppression factor to the second power,
$\langle\phi_{3}\rangle / M^u$ is assumed to be about $0.7$ \cite{deMedeirosVarzielas:2006fc}.
Note that if $\langle\phi_{3}\rangle / M^u$ was taken to be exactly or very close to $1$
then higher-order operators would be no-more suppressed than
lower-order operators, so such large expansion parameters are
of some concern, and in this respect the model is on a similar footing to that
\cite{deMedeirosVarzielas:2006fc}.

In order to avoid problems with unification,
we assume that the mass of the above messenger fields have masses close to
the conventional GUT scale, which suggests that the $\Delta_{27}$ family symmetry is also broken close to GUT scale.
In principle this could lead to GUT scale threshold effects, but we do not consider this further here.
Below the GUT scale the particle content of the model below the GUT scale is then the equivalent to that of the E$_6$SSM,
with the possible inclusion of additional SM singlet flavon fields close to the GUT scale
that will not affect the running of the gauge coupling constants.
As in the E$_6$SSM, below the GUT scale
there are three copies of complete $27$ representations of E$_6$
plus two additional electroweak doublets with masses of order the TeV scale, with the gauge coupling constants predicted to unify at the conventional GUT scale but with a
larger unified gauge coupling constant than in the MSSM \cite{King:2005jy}.

In section 2.5 we also discuss the messenger fields which are
responsible for the higher-order operators that generate the
effective $\mu$-term.  These particular messengers include
Higgs-like messengers and messengers that transform as singlets of
the Standard Model gauge group.

\begin{figure}
\center
\includegraphics[angle=0, scale=0.32]{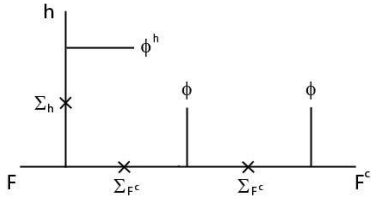} 
  \caption{\footnotesize This figure illustrates the type of messenger diagram that provides the dominant contribution to the Yukawa operators in Eq.\ref{eq:1}.}
   \label{b}
     \end{figure}

\subsubsection{The Effective Yukawa Matrices}

Inputting the above flavon VEVs into the superpotential given by
Eq.\ref{eq:1} generates the following effective Yukawa matrices
for the quarks and leptons:
\[ \mathbf{\lambda}^{ij}_u = \lambda_t \left( \begin{array}{ccc} 0 & \epsilon^3_u & \mathcal{O} (\epsilon^3_u)
\\ \epsilon^3_u & \epsilon^2_u & \mathcal{O} (\epsilon^2_u)
\\ \mathcal{O} (\epsilon^3_u) & \mathcal{O} (\epsilon^2_u) & 1 \end{array} \right) \hspace{1cm} \mathbf{\lambda}^{ij}_d = \lambda_t \left( \begin{array}{ccc} 0 & 1.5 \epsilon^3_d & 0.4 \epsilon^3_d
\\ 1.5 \epsilon^3_d & \epsilon^2_d & 1.3 \epsilon^2_d
\\ 0.4 \epsilon^3_d & 1.3 \epsilon^2_d & 1 \end{array} \right) \]
where $\epsilon_d = 3 \times \epsilon_u \approx 0.15$, and
$\lambda_t \approx 0.5 $ at the GUT scale \cite{Ross:2007az}.

The above form of Yukawa matrices have been shown to produce  a
realistic CKM matrix and realistic mass hierarchies for the up and
down quarks \cite{Ross:2007az}.
The $H_{45}$ Higgs field in the superpotential  Eq.\ref{eq:1} is
used to generate the Georgi-Jarlskog relations in the down quark
and charged lepton matrices so that the correct hierarchy in the
charged leptons is generated at the electroweak scale
\cite{Georgi:1979df}.

\subsection{Tri-Bi-Maximal Mixing}

To generate tri-bi-maximal mixing we use constrained sequential
dominance \cite{King:1998jw} in which the right-handed neutrinos have a hierarchy in
mass. This hierarchy is generated by the following operators:
\[W_{Maj} \sim \frac{1}{M} F^c_i F^c_j \theta^i \theta^j\]

\[\hspace{1.14cm} + \frac{1}{M^5} F^c_i F^c_j \phi^i_{23} \phi^j_{23} (\theta^k \overline{\phi}_{123 k} ) ( \theta^l \overline{\phi}_{3l}) \]

\begin{equation} \label{eq:WMaj}
\hspace{1.14cm} + \frac{1}{M^5} F^c_i F^c_j \phi^i_{123}
\phi^j_{123} (\theta^k \overline{\phi}_{123 k} ) ( \theta^l
\overline{\phi}_{123 l})
\end{equation}
where $\overline{\phi}_{123}$ and $\overline{\phi}_{3}$  are
triplets of $\Delta_{27}$, and $\theta$ is from a $\overline{27}$
of E$_6$ and a $\overline{3}$ of $\Delta_{27}$.

The above operators together with the dirac operator involving the
left-handed and right-handed neutrinos from the superpotential
$F_i F^{c}_j h_3$ create a conventional type I see-saw mechanism
which generates an effective Majorana mass matrix for the
left-handed neutrinos.  Due to form of the right-handed Majorana
mass matrix, and the dirac mass matrix from  Eq.\ref{eq:1}, the
effective Majorana matrix for the left-handed neutrinos is of a
form that is diagonalized by a tri-bi-maximal matrix (see
\cite{deMedeirosVarzielas:2006fc} for more details).

\subsection{The Effective $\mu$-Term and Inert Higgsino/Singlino Masses}

The $U(1)_N$ symmetry of the E$_6$SSM forbids any bilinear
superpotential terms for  the different Higgs fields.  Instead
effective bilinear terms come from the Pati-Salam superpotential
term  $\lambda^{ijk} S_i h_j h_k$ from Eq.\ref{eq:W},  where
$i,j,k = 1 \ldots 3$, once $S_3$ gets a VEV.  In terms of the
Standard Model gauge group, which is the symmetry of the model
discussed in this Letter, this superpotential term reduces to
$\lambda^{ijk} S_i h_{uj} h_{dk}$ where $h_i$ and $h_d$ denote up
and down Higgs fields. In this Letter we take the singlet fields
$S_i$, like the Higgs fields, to transform as a triplet of the
$\Delta_{27}$ symmetry.  We also take them to be odd under a
$Z^S_2$ symmetry and even under the $Z^h_2$ symmetry.  The
operator $\lambda^{ijk} S_i h_{uj} h_{dk}$ is thus forbidden and
is instead generated by the following higher-order operators:

\[
W_{\mu} \sim \frac{1}{M^3} S_i h_{uj} h_{dk}  (\phi^{S}_{3})^i
(\phi^{h}_{3})^j (\phi^{h}_{3})^k
\]

\[
\hspace{0.7cm} +\frac{1}{M^2} \epsilon^{jkl} S_i h_{uj} h_{dk}
(\phi^{S}_{3})^i (\overline{\phi}^h_{3})_l
\]

\begin{equation} \label{eq:Shh}
\hspace{0.7cm} +\frac{1}{M^2} \epsilon^{ijl} S_i h_{uj} h_{dk}
(\overline{\phi}^{S}_{3})_l (\phi^{h}_{3})^k +\frac{1}{M^2}
\epsilon^{ijl} S_i h_{uk} h_{dj}  (\overline{\phi}^{S}_{3})_l
(\phi^{h}_{3})^k
\end{equation}
where the flavons have the charges shown in Table 1 and we take the scale of
these flavon VEVs to be such that $\langle\phi^S_3\rangle / M_S = \epsilon_S$,
$\langle\overline{\phi}^S_3\rangle / M_h = \overline{\epsilon}_S$ and
$\langle\overline{\phi}^h_3\rangle / M_S = \overline{\epsilon}_h$ where $M_S$
is the mass scale of the singlet-like messengers and $M_h$ is the
mass scale of the Higgs-like messengers.
The messenger diagrams responsible for
generating the above higher-order operators are represented by
Figure 2.


The first operator in Eq.\ref{eq:Shh} is responsible for
generating an effective $\mu$-term for the third family of Higgs
fields once the flavons fields and the third family singlet field
$S_3$ obtain VEVs.  Since we assume that only the third family of
Higgs obtains a VEV, this effective $\mu$-term acts like the
$\mu$-term of the MSSM Higgs fields.  The effective $\mu$-term
will have a value $(0.8)^3 \langle S_3\rangle$, which will be approximately
$1~TeV$ if $\langle S_3\rangle = 2~TeV$, which is consistent with the
experimental bound for the mass of a $Z'$.  If we assume a
radiative symmetry breaking explanation for the third family
singlet field's VEV then it will be related to the singlet's soft
mass term.  This then explains the apparent requirement of the
$\mu$-term of the MSSM being related to the MSSM soft terms.

\begin{figure}
  \center
  \includegraphics[angle=0, scale=0.29]{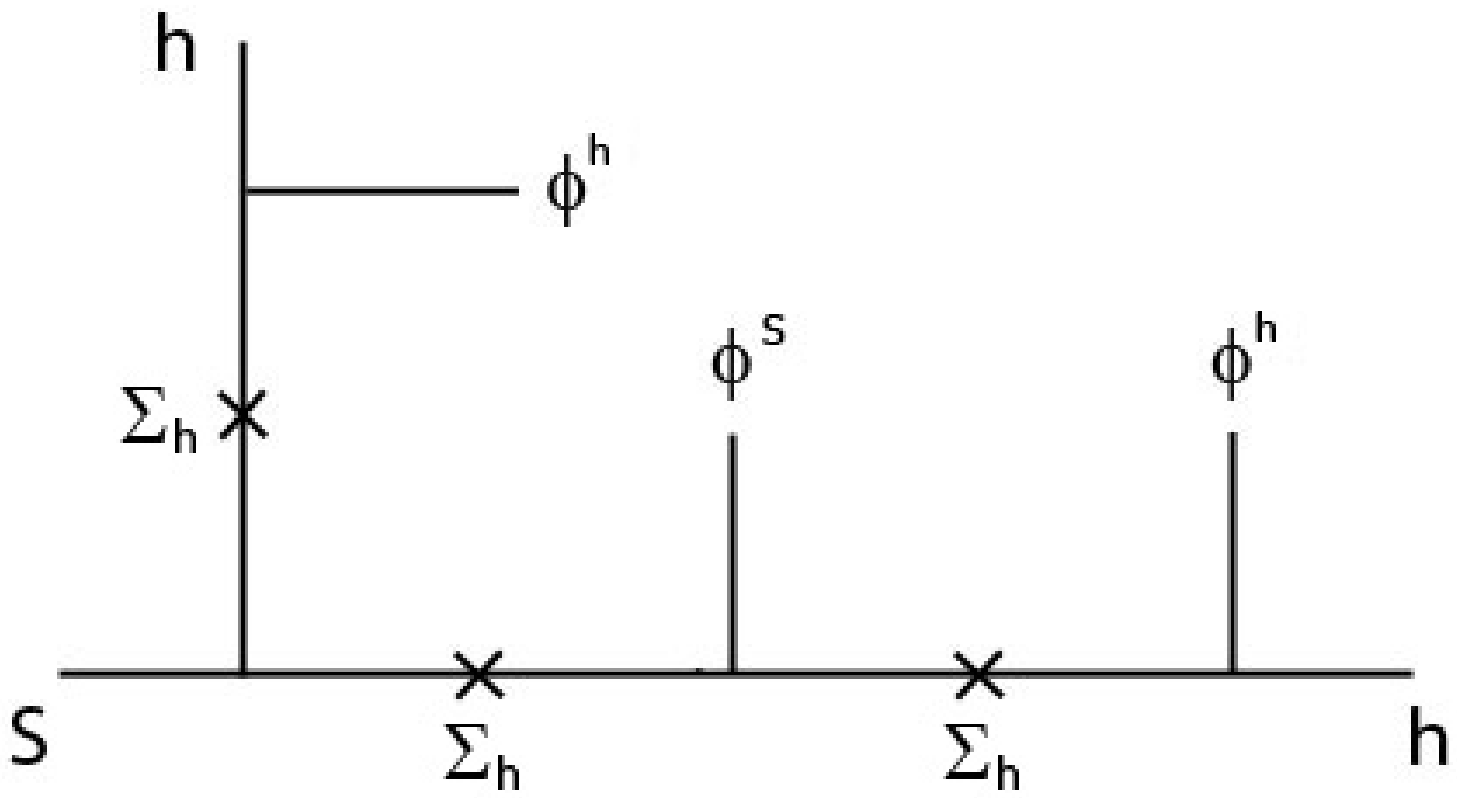} 
\hspace{0.5cm}
  \includegraphics[angle=0, scale=0.29]{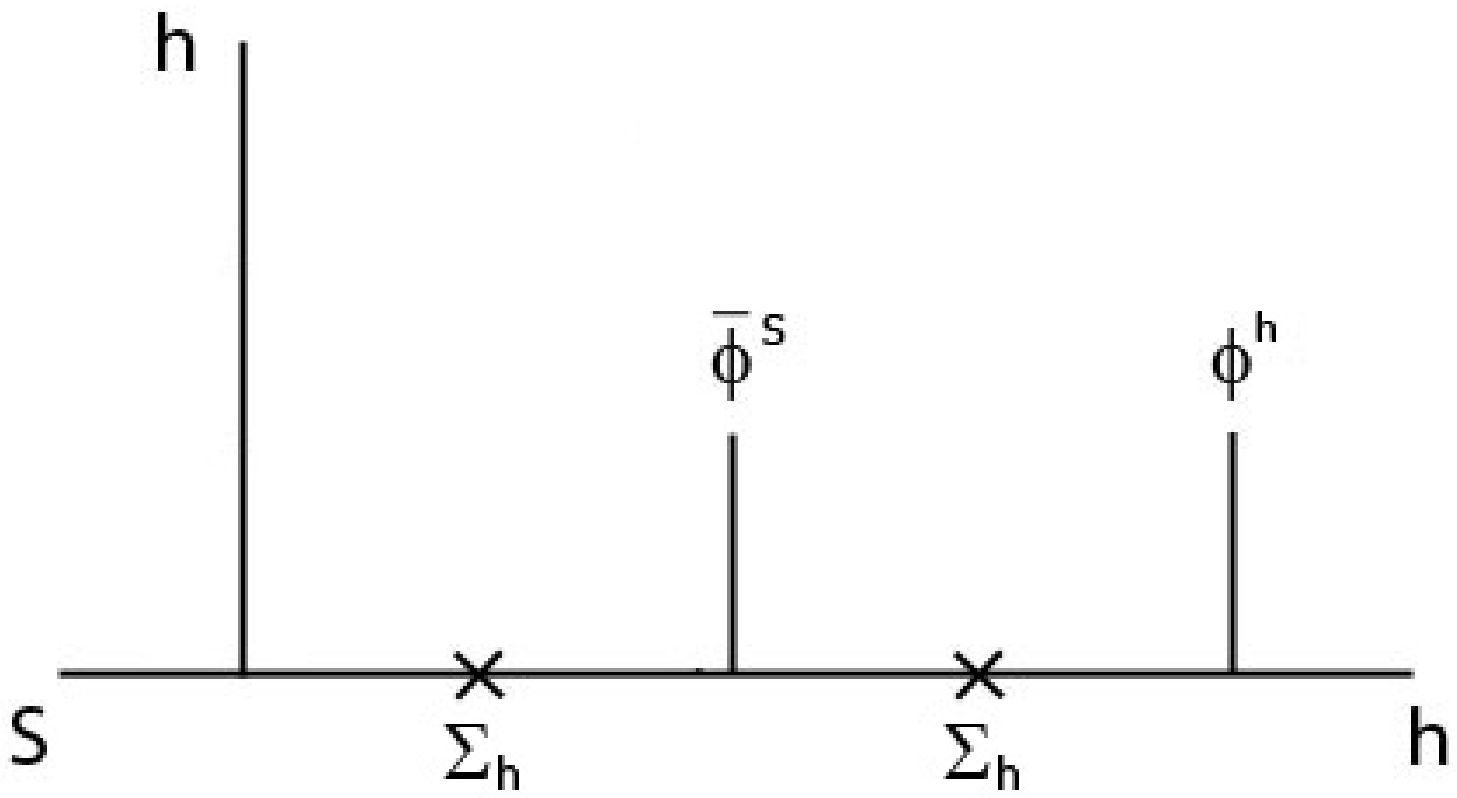}
  \hspace{0.5cm}
 \includegraphics[angle=0, scale=0.29]{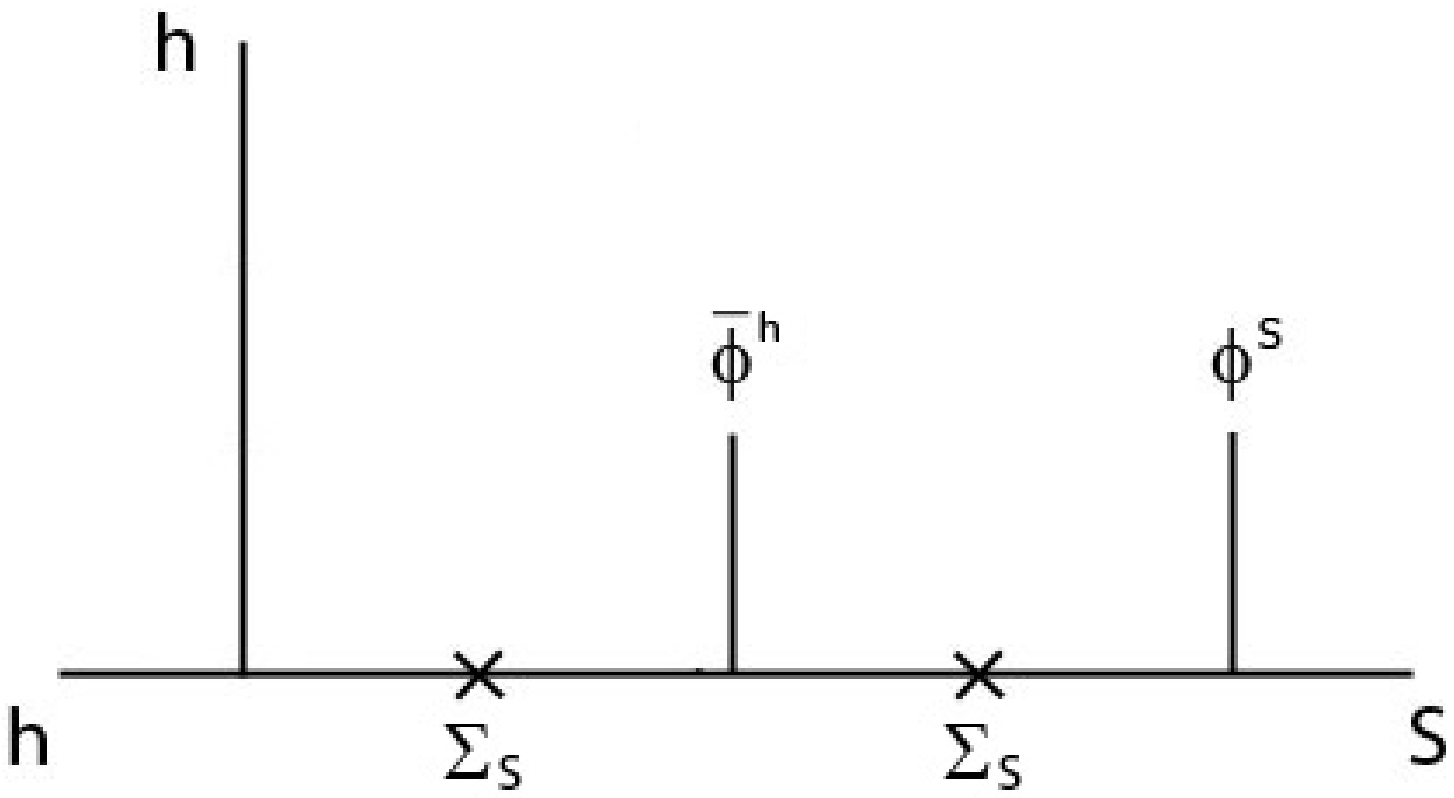} 
 \caption{\footnotesize This figure illustrates the type of messenger diagram that provides the dominant contribution to the effective $\mu$-term operators in Eq.\ref{eq:Shh}.}
   \label{b}
\end{figure}

The second and third operators in Eq.\ref{eq:Shh} are responsible
for providing mass to the first and second families of Higgsinos
and singlinos once the third family of Higgs fields and singlet
field obtain VEVs.  This results in a mixing between all of these
states which is represented by the following matrix:
\[ M^{inert} =  \left( \begin{array}{cc} A_{22} & A_{21}
\\ A_{21}^T & A_{11}  \end{array} \right) \]
This matrix is written in the basis $(\tilde{h}^0_{d2},
\tilde{h}^0_{u2}, \tilde{S}_2 | \tilde{h}^0_{d1},
\tilde{h}^0_{u1}, \tilde{S}_1)$ so that $A_{\alpha \beta}$ are $3
\times 3$ matrices where $\alpha, \beta = 1,2$.   Because of the
anti-symmetric tensor in the Eq.\ref{eq:Shh} we find that $A_{11}
= A_{22} = \mathbf{0}$, whereas $A_{21}$ is given by the
following:
\be
 A_{21} =  \left( \begin{array}{ccc} 0 & \epsilon_S \overline{\epsilon}_h \langle S^3\rangle & \overline{\epsilon}_S \langle h^3_u\rangle
\\ \epsilon_S \overline{\epsilon}_h \langle S^3\rangle& 0 & \overline{\epsilon}_S \langle h^3_d\rangle
\\ \overline{\epsilon}_S \langle h^3_u\rangle & \overline{\epsilon}_S \langle h^3_d\rangle & 0 \end{array} \right),
\ee where this matrix couples the states $(\tilde{h}^0_{d2},
\tilde{h}^0_{u2}, \tilde{S}_2)$ to the states $(\tilde{h}^0_{d1},
\tilde{h}^0_{u1}, \tilde{S}_1)$. In the limit of exact $Z^h_2$ and
$Z^S_2$ symmetry these Higgsino and singlino states will decouple
from the usual inert USSM states such as the third family of
Higgsinos, singlinos, wino and hypercharge bino fields.  See
\cite{Hall:2009aj} for a full discussion on the mixing between the
usual USSM states and the additional E$_6$SSM states where it is
also shown that the mixing between the $U(1)_N$ bino and Higgsino
and singlino fields is expected to be small.

The above Higgsino and singlino neutral states combine to form two
degenerate LSP states, approximately consisting of a Dirac state
formed from (dropping the tildes) $S_1$ and $S_2$, together with
two generally heavier approximately degenerate Dirac states formed
from ${h}^0_{d1}$ and ${h}^0_{u2}$ on the one hand and
${h}^0_{d2}$ and ${h}^0_{u1}$ on the other hand. With exact
R-parity the Dirac LSP state formed from $S_1$ and $S_2$ becomes a
dark matter candidate. However the masses of the degenerate LSPs
$S_1$ and $S_2$ can be split if the first and second generation of
Higgs and singlet fields are distinguished from one another.  One
way of achieving this is to  assume that the flavon field
$\phi^h_3$ gets small vacuum expectation values in its first and
second components of $\Delta_{27}$ such that $\langle\phi^h_3\rangle^T \propto
(\delta_1, \delta_2, 1)$ where $\delta_1,~\delta_2 \ll 1$.  This
might be expected to occur from higher-order operators that affect
the vacuum alignment of the fields.  Two WIMPs that are almost
degenerate in mass have been recently used to explain the DAMA
data \cite{Chang:2008gd}.


Note that although the $Z^h_2$ and $Z^S_2$ symmetries of this
model have combined to operate in a similar manner to the original
$Z^H_2$ symmetry of the E$_6$SSM, they allow fewer operators than
the latter.  The operators allowed by the original $Z^H_2$
symmetry but which are not present in this model are $S_3
h_{u\alpha} h_{d\alpha}$, $S_{\alpha} h_{u\alpha} h_{d3}$ and
$S_{\alpha} h_{u3} h_{d\alpha}$.  Such operators are responsible
for the $A_{22}$ and $A_{11}$ matrices being non-zero in the
E$_6$SSM.





\subsection{D-fermion Mass Terms}

For the $U(1)_N$ group of the E$_6$SSM to be anomaly free, all the
colour partners $D_i$ of the Higgs fields  from the three $27$
multiplets must have masses lower than the energy scale of
$\langle S_3\rangle$.  These masses come from the Pati-Salam superpotential
$\lambda^{ijk} S_i D_j D_k$ which is derived from the E$_6$
superpotential of the E$_6$SSM given by Eq.\ref{eq:W}.  In terms
of a Standard Model gauge symmetry we represent this operator as
$\lambda^{ijk} S_i D_j \overline{D}_k$ where $D$ is a triplet of
the strong force gauge group $SU(3)_c$ but $\overline{D}$ is an
anti-triplet.

In this Letter we assume that the exotic particles $D_i$, like the
Higgs fields transform as a triplet of $\Delta_{27}$ and have odd
$Z^h_2$ parity but even $Z^S_2$ parity.  The allowed higher order
operator thus mirrors the allowed operators that provide effective
$\mu$-terms for the Higgs fields:
\[
W_{D} \sim \frac{1}{M^3} S_i D_j \overline{D}_k  (\phi^{S}_{3})^i
(\phi^{h}_{3})^j (\phi^{h}_{3})^k
\]

\[
\hspace{0.8cm} +\frac{1}{M^2} \epsilon^{jkl} S_i D_j
\overline{D}_k  (\phi^{S}_{3})^i (\overline{\phi}^h_{3})_l
\]

\begin{equation} \label{eq:SDD}
\hspace{0.8cm} +\frac{1}{M^2} \epsilon^{ijl} S_i D_j
\overline{D}_k  (\overline{\phi}^{S}_{3})_l (\phi^{h}_{3})^k
+\frac{1}{M^2} \epsilon^{ijl} S_i D_k \overline{D}_j
(\overline{\phi}^{S}_{3})_l (\phi^{h}_{3})^k.
\end{equation}
The mass scale for the exotic-like messengers $\Sigma_{D,
\overline{D}}$ responsible for the operators in Eq.\ref{eq:SDD}
however need not be the same as the Higgs messengers.  We define
the messenger scales such that $M_D = M_{\overline{D}}$,
$\langle\phi^h_3\rangle / M_D \equiv \epsilon_D$, $\langle\overline{\phi}^h_3\rangle / M_S
\equiv \overline{\epsilon}_D$ and $\langle\overline{\phi}^S_3\rangle / M_D
\equiv \overline{\epsilon}'_S$. We also assume that the
exotic-like messengers, like the Higgs-like messengers only have
even $Z^h_2$ parity but can carry either even or odd $Z^S_2$
parity.   The messenger diagrams that are assumed to be
responsible for generating the higher-order operators in
Eq.\ref{eq:SDD} are analogous to those in Figure 2 but with the
Higgs fields and Higgs-like messenger fields replaced with exotic
fields and exotic-like messenger fields respectively.

The D-fermions thus obtain mass once the flavons and $S_3$
obtain an expectation value.  We write these masses in matrix form
$M^{Dij} D_i \overline{D}_j$ where $M^{Dij}$ is the following:
\[ M^{Dij} =  \left( \begin{array}{ccc} 0 & \epsilon_S \overline{\epsilon}_D & 0
\\ \epsilon_S \overline{\epsilon}_D & 0 & 0
\\ 0 & 0 & \epsilon_S \epsilon^2_D + \epsilon^3_D \end{array} \right) \langle S^3\rangle.\]
The parameters $\epsilon_S$, $\epsilon_D$ and
$\overline{\epsilon}_D$ can then be chosen for the exotic masses
to be larger than the experimental bound of  $300~GeV$.  Two of
the exotic states are predicted to be degenerate in mass with the
third also being degenerate in the approximation that
$\epsilon^2_D = \overline{\epsilon}_D$ and $\epsilon_D \ll
\epsilon_S$.  This mass structure is in stark contrast to the
hierarchical structure of the quarks and leptons despite all the
states being triplets of the family symmetry.

\subsection{D-fermion Decay and Proton Decay Suppression}

If the exotic $D$ particles are taken to have the same
$\Delta_{27}$, $Z^h_2$ and $Z^S_2$ quantum numbers as the Higgs
fields $h$, then they can
decay \cite{King:2005jy} via the following non-renormalizable operators:
\[
W_{Exotic} \sim \frac{1}{M^3} F^c_i F^c_j D_k   \phi^i_{3} \phi^j_{3}
(\phi^{h}_3)^{k}
\]

\[
\hspace{1.4cm} +\frac{1}{M^4} (F_i F_j + F^c_i F^c_j) D_k
\phi^i_{23} \phi^j_{23} (\phi^{h}_3)^{k} H_{45}
\]

\[
\hspace{1.4cm}+\frac{1}{M^3} (F_i F_j + F^c_i F^c_j) D_k
\phi^i_{123} \phi^j_{23} (\phi^{h}_3)^{k}
\]

\begin{equation} \label{eq:FFD}
\hspace{1.4cm}+\frac{1}{M^3} (F_i F_j + F^c_i F^c_j) D_k
(\phi^i_{123} \phi^j_3 + \phi^i_{3} \phi^j_{123}) (\phi^m_{123}
\overline{\phi}_{1m}) (\phi^{h}_3)^{k} H_{45}.
\end{equation}
However not all these operators can be allowed otherwise this would lead to
very rapid proton decay. Thus, we assume either the $Z^B_2$ or $Z^L_2$ discrete
symmetries that are used in the E$_6$SSM \cite{King:2005jy},
where these symmetries differentiate between different fermion $F,F^c$
components. Under
the $Z^B_2$ symmetry the leptons and D states are odd
whereas, under the $Z^L_2$ symmetry, only the leptons are odd and
all other particles are even.  These discrete symmetries
remove some of the above operators, such that the remaining operators
correspond to the D-states coupling as either diquarks or
leptoquarks. This effectively prevents
the proton from decaying due to couplings with the D-states,
but still allowing the latter states to decay, thus avoiding any
nucleosynthesis difficulties \cite{King:2005jy,Wolfram:1978gp}.
Note that this discrete symmetry breaks the Pati-Salam gauge
symmetry but respects the Standard Model gauge symmetry
assumed in this Letter.

In the limit that $\langle\phi^h_3\rangle^T \propto (0,0,1)$ exactly, the
decay channels of the exotic states in the model used in this
Letter will be different to those of the E$_6$SSM since only the
third generation of the exotic states couples directly to quarks
and leptons, whereas all three generations of the exotic states in
the E$_6$SSM interact directly with the quarks and leptons.  The
difference between the two models occurs because we have taken the
exotic states to transform under $Z^h_2$, which results in an
effective $Z^H_2$ symmetry for only the first and second
generation of exotic states.  In the E$_6$SSM however  all three
generations transform under $Z^H_2$.  This application of the
$Z^h_2$ symmetry results in the decay products of the first and
second generation of exotic states always involving a singlet
field $S_i$.

If instead $\langle\phi^h_3\rangle^T \propto (\delta_1,\delta_2,1)$ as
discussed in section 2.5, then all the $\Delta_{27}$ components of
the exotic states will mix via the mass terms presented in section
2.6.  This results in the same exotic decay channels as used in
the E$_6$SSM but with some being more suppressed since $\delta_1,
~\delta_2 \ll 1$.


\section{Summary}
In this Letter we have shown how
FCNC's due to models with three families of Higgs fields may be tamed
by the same family symmetry which predicts
tri-bi-maximal lepton mixing and provides a solution to the SUSY
FCNC and CP problems. To be concrete we have focussed on the
E$_6$SSM  where we have shown how the flavour problem
can be solved by using a $\Delta_{27}$ family symmetry.
The three 27 dimensional families of the E$_6$SSM, including the
three families of Higgs fields, transform in a triplet
representation of the $\Delta_{27}$ family symmetry, allowing the
family symmetry to commute with a possible high energy E$_6$
symmetry. The $\Delta_{27}$ family symmetry breaking considered
here, together with a vertical $Z^h_2$ symmetry which does not
distinguish between the three families, gives rise effectively to the $Z_2^H$
symmetry of the E$_6$SSM, which solves the flavour changing
neutral current problem of the three families of Higgs fields. The
main phenomenological predictions of the model are tri-bi-maximal
mixing for leptons, two almost degenerate LSPs and two almost
degenerate families of colour triplet D-fermions, providing a
clear prediction for the LHC. In addition the D-term vacuum alignment
mechanism described here leads to PGBs with masses below the TeV scale, and possibly much lighter,
which appears to be a quite general and robust prediction of any model based on the
D-term vacuum alignment mechanism.

\textbf{Acknowledgments}
We would like to thank C.Luhn and G.Ross for useful
discussions about radiative symmetry breaking.











\begin{thebibliography}{99}

\bibitem{Chung:2003fi}
For a recent review see e.g. D.~J.~H.~Chung, L.~L.~Everett,
G.~L.~Kane, S.~F.~King, J.~Lykken, L.~T.~Wang,
  Phys. \ Rept. \  {\bf 407} (2005) 1.

\bibitem{BasteroGil:2000bw}
  M.~Bastero-Gil, C.~Hugonie, S.~F.~King, D.~P.~Roy and S.~Vempati,
  Phys.\ Lett.\  B {\bf 489} (2000) 359
  [arXiv:hep-ph/0006198].

\bibitem{Fayet:1977yc}
  P.~Fayet,
  Phys.\ Lett.\  B {\bf 69} (1977) 489.

\bibitem{Kalinowski:2008iq}
  J.~Kalinowski, S.~F.~King and J.~P.~Roberts,
  JHEP {\bf 0901} (2009) 066
  [arXiv:0811.2204 [hep-ph]].




\bibitem{King:2005jy}
  S.~F.~King, S.~Moretti and R.~Nevzorov,
  Phys.\ Rev.\  D {\bf 73} (2006) 035009
  [arXiv:hep-ph/0510419];
  S.~F.~King, S.~Moretti and R.~Nevzorov,
  Phys.\ Lett.\  B {\bf 634} (2006) 278
  [arXiv:hep-ph/0511256];
  S.~F.~King, S.~Moretti and R.~Nevzorov,
  Phys.\ Lett.\  B {\bf 650} (2007) 57
  [arXiv:hep-ph/0701064];
  P.~Athron, S.~F.~King, D.~J.~.~Miller, S.~Moretti, R.~Nevzorov and R.~Nevzorov,
  arXiv:0901.1192 [hep-ph];
  P.~Athron, S.~F.~King, D.~J.~Miller, S.~Moretti and R.~Nevzorov,
  arXiv:0904.2169 [hep-ph].


\bibitem{Langacker:2008yv}
  P.~Langacker,
  arXiv:0801.1345 [hep-ph].


\bibitem{Howl:2007zi}
  R.~Howl and S.~F.~King,
  JHEP {\bf 0801} (2008) 030
  [arXiv:0708.1451 [hep-ph]];
  R.~Howl and S.~F.~King,
  Phys.\ Lett.\  B {\bf 652} (2007) 331
  [arXiv:0705.0301 [hep-ph]].


\bibitem{HPS}
P.~F.~Harrison, D.~H.~Perkins and W.~G.~Scott,
Phys.\ Lett.\ B {\bf 530} (2002) 167 [arXiv:hep-ph/0202074].


\bibitem{King:2009ap}
  S.~F.~King and C.~Luhn,
  arXiv:0908.1897 [hep-ph].






\bibitem{Lam:2009hn}
  C.~S.~Lam,
  arXiv:0907.2206 [hep-ph].



\bibitem{Ma:2007wu}
  E.~Ma and G.~Rajasekaran,
  Phys.\ Rev.\  D {\bf 64} (2001) 113012
  [arXiv:hep-ph/0106291].








\bibitem{Altarelli:2006kg}
  G.~Altarelli,
  arXiv:hep-ph/0611117;
G.~Altarelli, F.~Feruglio and Y.~Lin,
  Nucl.\ Phys.\  B {\bf 775} (2007) 31
  [arXiv:hep-ph/0610165];
  G.~Altarelli and F.~Feruglio,
  Nucl.\ Phys.\  B {\bf 741} (2006) 215
  [arXiv:hep-ph/0512103];
  G.~Altarelli and F.~Feruglio,
  Nucl.\ Phys.\  B {\bf 720} (2005) 64
  [arXiv:hep-ph/0504165];
  M.~C.~Chen and S.~F.~King,
  JHEP {\bf 0906} (2009) 072
  [arXiv:0903.0125 [hep-ph]];
  T.~J.~Burrows and S.~F.~King,
  arXiv:0909.1433 [hep-ph].



\bibitem{King:2009mk}
  S.~F.~King and C.~Luhn,
  Nucl.\ Phys.\  B {\bf 820} (2009) 269
  [arXiv:0905.1686 [hep-ph]].

\bibitem{King:2006np}
  S.~F.~King and M.~Malinsky,
  Phys.\ Lett.\  B {\bf 645} (2007) 351
  [arXiv:hep-ph/0610250].


\bibitem{deMedeirosVarzielas:2006fc}
  I.~de Medeiros Varzielas, S.~F.~King and G.~G.~Ross,
  Phys.\ Lett.\  B {\bf 648} (2007) 201
  [arXiv:hep-ph/0607045].







\bibitem{King:2006me}
  S.~F.~King and M.~Malinsky,
  JHEP {\bf 0611} (2006) 071
  [arXiv:hep-ph/0608021];
S.~F.~King,
JHEP {\bf 0508} (2005) 105;
S.~Antusch and S.~F.~King,
  Nucl.\ Phys.\  B {\bf 705} (2005) 239
  [arXiv:hep-ph/0402121].


\bibitem{King:2003rf}
  S.~F.~King and G.~G.~Ross,
  Phys.\ Lett.\  B {\bf 520} (2001) 243
  [arXiv:hep-ph/0108112];
  S.~F.~King and G.~G.~Ross,
  Phys.\ Lett.\  B {\bf 574} (2003) 239
  [arXiv:hep-ph/0307190];
  I.~de Medeiros Varzielas and G.~G.~Ross,
  Nucl.\ Phys.\  B {\bf 733} (2006) 31
  [arXiv:hep-ph/0507176].

















\bibitem{deMedeirosVarzielas:2005qg}
  I.~de Medeiros Varzielas, S.~F.~King and G.~G.~Ross,
  Phys.\ Lett.\  B {\bf 644} (2007) 153
  [arXiv:hep-ph/0512313].


















\bibitem{Antusch:2008jf}
  S.~Antusch, S.~F.~King and M.~Malinsky,
  JHEP {\bf 0806} (2008) 068
  [arXiv:0708.1282 [hep-ph]];
  S.~Antusch, S.~F.~King, M.~Malinsky and G.~G.~Ross,
  Phys.\ Lett.\  B {\bf 670} (2009) 383
  [arXiv:0807.5047 [hep-ph]].


\bibitem{Howl:2008xz}
  R.~Howl and S.~F.~King,
  JHEP {\bf 0805} (2008) 008
  [arXiv:0802.1909 [hep-ph]].


\bibitem{Greene:1986jb}
  B.~R.~Greene, K.~H.~Kirklin, P.~J.~Miron and G.~G.~Ross,
  Nucl.\ Phys.\  B {\bf 292} (1987) 606;
  I.~de Medeiros Varzielas and G.~G.~Ross,
  arXiv:hep-ph/0612220;

\bibitem{Ross:2007az}
  G.~Ross and M.~Serna,
  arXiv:0704.1248 [hep-ph].

\bibitem{Georgi:1979df}
  H.~Georgi and C.~Jarlskog,
  Phys.\ Lett.\  B {\bf 86} (1979) 297.

\bibitem{King:1998jw}
S.~F.~King,
Phys.\ Lett.\ B {\bf 439} (1998) 350 [arXiv:hep-ph/9806440];
S.~F.~King,
Nucl.\ Phys.\ B {\bf 562} (1999) 57 [arXiv:hep-ph/9904210];
S.~F.~King,
Nucl.\ Phys.\ B {\bf 576} (2000) 85 [arXiv:hep-ph/9912492];
S.~F.~King,
JHEP {\bf 0209} (2002) 011 [arXiv:hep-ph/0204360];
  S.~Antusch and S.~F.~King,
  New J.\ Phys.\  {\bf 6} (2004) 110
  [arXiv:hep-ph/0405272];
  S.~F.~King,
  JHEP {\bf 0508}, 105 (2005)
  [arXiv:hep-ph/0506297].


\bibitem{Hall:2009aj}
  J.~P.~Hall and S.~F.~King,
  arXiv:0905.2696 [hep-ph].


\bibitem{Chang:2008gd}
  S.~Chang, G.~D.~Kribs, D.~Tucker-Smith and N.~Weiner,
  Phys.\ Rev.\  D {\bf 79} (2009) 043513
  [arXiv:0807.2250 [hep-ph]].


\bibitem{Wolfram:1978gp}
  S.~Wolfram,
  Phys.\ Lett.\  B {\bf 82} (1979) 65.
  C.~B.~Dover, T.~K.~Gaisser and G.~Steigman,
  Phys.\ Rev.\ Lett.\  {\bf 42} (1979) 1117.
  T.~K.~Hemmick {\it et al.},
  Phys.\ Rev.\  D {\bf 41} (1990) 2074.






\end{thebibliography}
\end{document}